\documentclass[twoside,12pt]{article}
\usepackage{natbib}
\usepackage{geometry}
\usepackage{amssymb,amsthm,amscd,amsfonts,amsmath}
\usepackage[pdftex]{graphicx}
\usepackage[pdftex]{color}
\usepackage{latexsym}  
\usepackage[T1]{fontenc}
\usepackage{times}
\usepackage[T1,mtbold]{mathtime}
\usepackage{mathrsfs}  
\usepackage{bbm}  
\usepackage{multirow}
\usepackage[small,compact]{titlesec}

\bibpunct[; ]{(}{)}{;}{a}{}{,}
\bibdata{../refs/braun_refs}

\newtheorem{theorem}{Theorem}

\newtheorem{claim}[theorem]{Claim}

\newtheorem{lemma}[theorem]{Lemma}

\newcommand{\mysec}[1]{Section~\ref{sec:#1}}
\newcommand{\myapx}[1]{Appendix~\ref{apx:#1}}

\newcommand{\myeqref}[1]{\eqref{eq:#1}}
\newcommand{\myfig}[1]{\ref{fig:#1}}
\newcommand{\mytbl}[1]{\ref{tbl:#1}}

\renewcommand{\:}{\text{:}}
\newcommand{\defsym}{\ \:\!\!=}
\newcommand{\E}{\mathbbmss{E}}
\newcommand{\defeq}{\mathrel{\mathop:}=}
\newcommand{\g}{\ |\ }
\newcommand{\dist}{\ \sim\ }
\newcommand{\inddist}{\ \stackrel{\text{ind}}{\sim}\ }
\newcommand{\iid}{\ \stackrel{\text{iid}}{\sim}\ }
\newcommand{\pr}[1]{p\!\left(#1\right)}
\newcommand{\qr}[1]{q\!\left(#1\right)}
\newcommand{\D}{\mathcal{D}}

\newcommand{\KLqlp}{\text{KL}\left[q_\lambda\,||\,p\right]}
\newcommand{\argmin}{\operatornamewithlimits{argmin}}
\newcommand{\argmax}{\operatornamewithlimits{argmax}}
\DeclareMathOperator{\diag}{diag}
\DeclareMathOperator{\tr}{tr}
\DeclareMathOperator{\vecop}{vec}
\newcommand{\mrd}{\mathrm{d}}
\newcommand{\cov}[1]{\text{Cov}(#1)}

\newcommand{\hess}[2]{\frac{\partial #1}{\partial #2 \partial #2^\top}}

\newcommand{\tmL}{\tilde{\mathcal{L}}}

\usepackage[doublespacing]{setspace}

\begin{document}

\begin{singlespace}

\title{Variational inference for large-scale models of discrete choice}
\author{Michael Braun \\
MIT Sloan School of Management\\
Massachusetts Institute of Technology\\
Cambridge, MA 02142 \and Jon McAuliffe \\
Statistics Department \\
University of Pennsylvania\\
Wharton School\\
Philadelphia, PA 19104}
\date{6 January 2008\thanks{%
The authors gratefully acknowledge research assistance from Sandy Spicer and
Liz Theurer.}}

\maketitle

\vspace{-1cm}

\begin{abstract}
  Discrete choice models are commonly used by applied statisticians in
  numerous fields, such as marketing, economics, finance, and operations
  research. When agents in discrete choice models are assumed to have
  differing preferences, exact inference is often intractable. Markov chain
  Monte Carlo techniques make approximate inference possible, but the
  computational cost is prohibitive on the large data sets now becoming
  routinely available. Variational methods provide a deterministic
  alternative for approximation of the posterior distribution. We derive
  variational procedures for empirical Bayes and fully Bayesian inference
  in the mixed multinomial logit model of discrete choice. The algorithms
  require only that we solve a sequence of unconstrained optimization
  problems, which are shown to be convex. Extensive simulations demonstrate
  that variational methods achieve accuracy competitive with Markov chain
  Monte Carlo, at a small fraction of the computational cost. Thus,
  variational methods permit inferences on data sets that otherwise could
  not be analyzed without bias-inducing modifications to the underlying
  model.
\end{abstract}

\end{singlespace}

\section{Introduction}

Discrete choice models have a long history in statistical analysis,
appearing in applications as varied as the analysis of consumer choice data
\citep{GuadagniLittle83,FaderHardie96}, transportation planning
\citep{Theil69,McFadden74,LermanBenAkiva85}, economic demand estimation
\citep{TrainMcFadden87,ReveltTrain98}, new product development
\citep{MooreLouviere99}, portfolio analysis \citep{UhlerCragg71} and health
services deployment \citep{HallKennyEtAl2002}. They apply to situations
where agents (also called choosers or decision-makers) select items from a
finite collection of alternatives (the choice set), either once or
repeatedly over time. For example, in a marketing context, agents are
``households''; each household makes a number of ``trips'' to a store, and
we observe the items selected for purchase on each trip.

Heterogeneous discrete choice models, which allow preferences to differ
across agents, are based on a hierarchical regression formulation. We have
agents numbered $h = 1, \ldots, H$, each with an unseen parameter vector
$\theta_h$ encoding preferences over item attributes. We observe one or
more choice events $Y_h \sim p(y_h \g \theta_h)$ per agent. The
$\theta_h$'s are modeled as independent draws from a prior distribution
$p(\theta_h \g \phi)$, where $\phi$ is a hyperparameter. This prior
represents the heterogeneity of preferences across the population.
Inference in such a hierarchical model allows us to pool information across
decision-makers. If we use an empirical Bayes point estimate of
$\phi$~\citep{Robbins55}, the posterior distribution of each
$\theta_h$ depends on all of $Y_1, \ldots, Y_H$, through the common
estimate $\hat{\phi}$. In a fully Bayesian setup, integrating out the
random variable $\phi$ creates similar dependence.

The marginal likelihood corresponding to one agent in a heterogeneous model
is
\begin{equation}
  p(y_h \g \phi) = \int \! p(y_h \g \theta_h) p(\theta_h \g \phi)
  \ d \theta_h \ .\label{eq:marginal}
\end{equation}
In most cases, including the ``random utility'' discrete choice model we
study in this paper,~\myeqref{marginal} does not exist in closed form. As a
consequence, we must use approximate methods both for empirical Bayes and
fully Bayesian inference.

A standard empirical Bayes technique is to approximate~\myeqref{marginal}
using Monte Carlo integration. But to match the asymptotics of maximum
likelihood, the number of draws per agent must grow faster than the square
root of the number of agents~\citep{Train2003}, which is infeasible for
large-scale problems. The usual approach to the fully Bayesian random
utility model is Markov chain Monte Carlo (MCMC)
simulation~\citep{AlbertChib93,AllenbyLenk94,AllenbyRossi99}. MCMC provides
approximate draws from the joint posterior distribution on $\theta_1,
\ldots, \theta_H$ and $\phi$. The draws enable the estimation
of~\myeqref{marginal} and related integrals. However, the more agents there
are in the data set, the more MCMC output we need to collect and store --
even if we are only interested in $\phi$, we still need repeated draws for
all of $\theta_1, \ldots, \theta_H$.

Variational methods~\citep{JordanEtAl1999,WainwrightJordan2003} offer a
deterministic alternative for approximate inference. With variational
inference, we maximize a data-dependent lower bound on the marginal
likelihood~\myeqref{marginal}, over a set of auxiliary parameters distinct
from the model parameters. In the fully Bayesian specification, the end
result is an approximate joint posterior distribution for $\theta_1,
\ldots, \theta_H$ and $\phi$. For empirical Bayes, variational techniques
lead to a point estimate $\hat{\phi}$ as well as an approximate posterior
distribution for the $\theta_h$'s.

The main advantage of variational methods versus MCMC is computational
efficiency. Variational inference algorithms typically converge to their
final approximation in far less time than it takes to generate an adequate
number of MCMC draws. This advantage comes at the cost of a biased
approximation, in contrast to the consistency guarantees that accompany
MCMC. We give evidence in \mysec{results} that, for our random utility
discrete choice model, variational bias is negligible, and the
computational speedup is very large. Variational convergence is also easy
to assess, in contrast to MCMC.

Furthermore, the size of the variational representation is fixed, while the
size of the MCMC approximation increases with the number of draws.
Variational techniques can therefore be applied to much larger data sets.
For example, with 10,000 decision-makers and 20,000 total MCMC draws, using
a 25-dimensional $\theta_h$ and corresponding $\phi$, the MCMC
representation of the posterior exceeds 2~GB---if we discard half the draws
for burn-in of the chain. In fact, many data sets today contain
observations on millions of agents, models can contain far more than 25
preference parameters, and MCMC chains may require hundreds of thousands of
iterations.

These difficulties are well known. MCMC is rarely applied to large-scale
heterogeneous models. Indeed, to address scalability, it is common to work
with data from a subset of individuals, or a subset of choice items.
However, this approach discards information that is valuable in the
inferential process, and it can lead to biased
estimates~\citep{BradlowZanutto2006}.

In this paper, we derive variational algorithms for a common discrete
choice model -- the mixed multinomial logit (MML) model. We study this
model because it the workhorse of discrete choice theory and is well known
in many disciplines, including economics and marketing. There are other
popular discrete choice models in the literature, but the mixed multinomial
logit has appeal: it is conceptually simple, yet still exhibits the MCMC
inference issues just described.

The rest of the paper is organized as follows. \mysec{mml} presents the
details of the MML model. In \mysec{vimml}, we describe variational
procedures suitable for empirical Bayes and fully Bayesian inference under
the MML model. These procedures include a novel application of the delta
method for moments to variational inference. In \mysec{results}, we compare
variational methods to MCMC for the MML model, using an extensive suite of
simulated data sets. \mysec{discussion} closes with discussion and future
directions. Technical arguments and derivations are relegated to
Appendixes~\ref{apx:varinf},~\ref{apx:delta-method},~and~\ref{apx:convexity}.

\section{The mixed multinomial logit model of discrete choice}
\label{sec:mml}

Let there be $H$ agents, indexed $h = 1, \ldots, H$. We observe a total of
$T_h$ choice-event outcomes for agent $h$. At each choice event, the agent
selects from among a fixed set of $J$ items, indexed $j = 1, \ldots, J$.
The items are differentiated according to $K$ attributes, indexed $k = 1,
\ldots, K$. The $j$th item's value for the $k$th attribute can vary across
agents and from one choice event to another. For example, households might
shop at different stores charging various prices for the same good, and the
price of a good may change over time within a single store. We denote by
$x_{ht}$ the $J \times K$ matrix of attribute values, also called
covariates, that agent $h$ encounters at her $t$th choice event. The $j$th
row of $x_{ht}$ is denoted $x_{htj}^\top$. The outcome of this choice event
is the observed categorical random variable $y_{ht}$, which we represent as
a $J \times 1$ indicator vector.

We use the observed $(x_{ht}, y_{ht})$ pairs to infer which attributes have
the strongest association with item choice. To this end, let $U_{htj}$
denote the utility that accrues to agent $h$ if she chooses item $j$ at her
$t$th choice event. This approach, called a ``random utility
model''~\citep{Train2003}, assumes utility is a noisy linear function of
the attributes: $U_{htj} = \beta_h^\top x_{htj} + e_{htj}$. Here, $\beta_h$
is a $K \times 1$ vector of agent-specific ``tastes'' or ``preference
loadings'' for the item attributes, and $e_{htj}$ is a random error term
representing unobserved utility.

We assume that each agent, at each choice event, selects the item
maximizing her utility. In the mixed multinomial logit model, we further
assume the random error terms $e_{htj}$ are iid from a Gumbel Type 2
distribution. The implied choice probabilities turn out to be
\begin{equation}
  P(y_{htj} = 1 \g x_{ht}, \beta_h) =
  \frac{ \exp( \beta_h^\top x_{htj} ) }
  { \sum_{j^\prime} \exp( \beta_h^\top x_{htj^\prime} ) } \ , \quad
  j = 1, \ldots, J \label{eq:mnl}
\end{equation}
\citep{McFadden74}. In discrete choice modeling, the right-hand side
of~\myeqref{mnl} is called the ``multinomial logit'' distribution, denoted
$\text{MNL}(x_{ht}, \beta_h)$. It is essentially the same as the
multi-logistic function used in polychotomous logistic regression, and it
is often called the soft-max function in machine learning research.

We further assume that $\beta_{1\:H}$ are iid from a \mbox{$K$-variate}
normal distribution with mean vector $\zeta$ and covariance matrix
$\Omega$, which we write as $\mathcal{N}_K(\zeta, \Omega)$. For empirical
Bayes estimation, the model is now completely specified:
\begin{alignat}{3}
  y_{ht} & \g x_{ht}, \beta_h &&
  \inddist \text{MNL}\!\left(x_{ht}, \beta_h\right), &\qquad&
   h = 1,\ldots,H , \quad t = 1, \ldots, T_h , \label{eq:yht} \\
  \beta_h & \g \zeta, \Omega &&
  \iid \mathcal{N}_K\!\left(\zeta, \Omega\right), && h = 1,\ldots,H \ .
  \label{eq:betah}
\end{alignat}
The top-level parameters $\zeta$ and $\Omega$, to be estimated by maximum
marginal likelihood, represent the distribution of attribute preferences
across the population. In particular, $\Omega$ gives us information about
the correlation of preferences between agents.

A fully Bayesian approach requires hyperprior distributions for $\zeta$ and
$\Omega$. As is standard, we use conditionally conjugate distributions:
\begin{align}
  \zeta \g \beta_0, \Omega_0 &
  \dist \mathcal{N}_K\!\left(\beta_0, \Omega_0\right), &
  \Omega \g S, \nu & \dist \mathcal{W}^{-1}(S^{-1}, \nu). \label{eq:zeta-iw}
\end{align}
In~\myeqref{zeta-iw}, $\beta_0$ and $\Omega_0$ are pre-specified
hyperparameters; $\mathcal{W}^{-1}(S^{-1}, \nu)$ is the inverse Wishart
distribution with scale matrix $S^{-1}$ and $\nu$ degrees of freedom; and
$S$ and $\nu$ are hyperparameters fixed in advance. We call the fully
Bayesian approach to MML model inference ``hierarchical Bayes.''

\section{Variational inference for the MML model} \label{sec:vimml}

We have presented the component hierarchical distributions in the mixed
multinomial logit model. Now we turn to the question of estimation and
inference procedures. In the following, variable names inside of $p(\cdot)$
are used to distinguish among densities: we denote the pdfs in
\myeqref{yht}--\myeqref{zeta-iw} by $\pr{y_{ht} \g x_{ht}, \beta_h}$,
$\pr{\beta_h \g \zeta, \Omega}$, $\pr{\zeta \g \beta_0, \Omega_0}$, and
$\pr{\Omega \g S, \nu}$, respectively. We let $\D = \{ \, x_{ht},
y_{ht} \}$ denote all observed variables, i.e., the data.

For the empirical Bayes version of the MML model, the posterior density of
the latent preference vectors, $\pr{\beta_{1\:H} \g \D, \zeta, \Omega}$, is
\begin{equation}
  \prod_{h=1}^H
  \frac{
    \pr{\beta_h \g \zeta, \Omega}
     \prod_{t=1}^{T_h}  \pr{y_{ht} \g x_{ht}, \beta_h}
  }{
    \int \! \pr{\beta_h \g \zeta, \Omega}
    \prod_{t=1}^{T_h} \pr{y_{ht} \g x_{ht}, \beta_h} \, d \beta_h
  } \ .
  \label{eq:ebpost}
\end{equation}
The joint posterior density for hierarchical Bayes, $\pr{ \beta_{1\:H} ,
  \zeta , \Omega \g \D }$, is
\begin{equation}
  \frac{ \pr{\zeta} \pr{\Omega}
    \ \prod_{h=1}^H \pr{\beta_h \g \zeta, \Omega}
    \ \prod_{t=1}^{T_h} \pr{y_{ht} \g x_{ht}, \beta_h} }
  { \int \pr{\zeta} \pr{\Omega}
    \ \prod_{h=1}^H \, \int \! \pr{\beta_h \g \zeta, \Omega}
    \ \prod_{t=1}^{T_h} \pr{y_{ht} \g x_{ht}, \beta_h}
    \, d \beta_h \, d \zeta \, d \Omega } \ .
  \label{eq:hbpost}
\end{equation}
The numerator in both cases is the joint density of latent and observed
variables, computed by multiplying together the densities defined in the
model hierarchy \myeqref{yht}--\myeqref{zeta-iw}.

The integrals appearing in these posterior densities have no closed form.
As a consequence, exact inference is intractable. Variational inference is
a deterministic alternative to the MCMC methods usually applied to this
problem~\citep{RossiAllenbyMcCulloch2005}. A variational algorithm selects
from a pre-specified family of distributions $\mathcal{Q}$ the best
approximation to the true posterior distribution. We define $\mathcal{Q}$
so that all of its members permit tractable probability calculations. Then,
wherever we need the true posterior, such as for an expectation, we use the
approximating variational distribution instead. This plug-in idea underlies
MCMC methods as well -- in place of the true posterior, we substitute the
empirical distribution of the MCMC posterior draws.

\subsection{Variational empirical Bayes}

In this section we give an overview of a variational algorithm for
approximate empirical Bayes estimation in the MML model. \myapx{varinf}
fills in the details. We first specify a family of approximating
distributions $\mathcal{Q} \defsym \{ \qr{\beta_{1\:H} \g \lambda}\ \:\
\lambda \in \Lambda \}$ for the true posterior
distribution~\myeqref{ebpost}. Since this posterior factors over $h$, we
take $\mathcal{Q}$ to be a family of factored distributions as well, so
that $\qr{\beta_{1\:H} \g \lambda} \defsym \prod_h \qr{\beta_h \g
  \lambda_h}$. In particular, each factor $\qr{\beta_h \g \lambda_h}$ is a
$K$-variate normal density, with mean $\mu_h$ and covariance matrix
$\Sigma_h$.

For the particular data set at hand, we want to find $\qr{\beta_{1\:H} \g
  \lambda^*}$, the best approximation in $\mathcal{Q}$ to the posterior
distribution~$\pr{\beta_{1\:H} \g \D, \zeta, \Omega}$. To make the idea of
a best approximation precise, we measure discrepancy with the
Kullback-Leibler (KL) divergence (also called the relative entropy).
Shortening $\qr{\beta_{1\:H} \g \lambda}$ to $q_\lambda$, the optimal
variational parameters are given by
\begin{equation}
  \lambda^* = \argmin_{\lambda \in \Lambda} \ \KLqlp . \label{eq:varinf}
\end{equation}
We can express the KL divergence between $q_\lambda$ and $p$ as
\begin{align}
  \KLqlp & = \E_{q_\lambda} \log\left[ \frac{\qr{\beta_{1\:H} \g \lambda}}
    {\pr{\beta_{1\:H} \g \D, \zeta, \Omega}} \right]
  \label{eq:kldecomp1} \\
  & = -\mathcal{L}(\lambda; \zeta, \Omega) + \log\pr{\D \g \zeta,
    \Omega} \label{eq:kldecomp2}
\end{align}
where we define
\begin{equation}
  \mathcal{L}(\lambda; \zeta, \Omega) \defsym - \E_{q_\lambda} \log\left[
    \frac{\qr{\beta_{1\:H} \g \lambda}}
    {\pr{\beta_{1\:H},\D \g \zeta, \Omega}} \right] .
  \label{eq:elbo}
\end{equation}
Here, $\E_{q_\lambda}$ denotes an average over $\beta_{1\:H}$, using $\qr{
  \beta_{1\:H} \g \lambda}$. Because $\KLqlp$ is
non-negative,~\myeqref{kldecomp2} implies that, for all distributions
$q_\lambda$,
\begin{equation}
  \mathcal{L}(\lambda; \zeta, \Omega) \leq \log p(\D \g \zeta, \Omega) \ .
  \label{eq:elbo2}
\end{equation}
The likelihood is called the ``evidence'' in some contexts; we therefore
christen $\mathcal{L}(\lambda; \zeta, \Omega)$ the \textit{evidence lower
  bound (ELBO)} function.

Using~\myeqref{kldecomp2}, we can formulate a maximization problem
equivalent to~\myeqref{varinf}, having the same optimal point
$\qr{\beta_{1\:H} \g \lambda^*}$:
\begin{equation}
  \lambda^* = \argmax_{\lambda \in \Lambda}
  \ \mathcal{L}(\lambda; \zeta, \Omega) \ .
  \label{eq:elboopt1}
\end{equation}
The equivalence of~\myeqref{varinf} and~\myeqref{elboopt1}, together with
the bound~\myeqref{elbo2}, shows that the best approximation in
$\mathcal{Q}$ to the posterior yields the tightest lower bound on the
marginal likelihood.

We work with~\myeqref{elboopt1} rather than~\myeqref{varinf} -- to evaluate
the KL divergence, we would need to compute the marginal likelihood,
bringing us back to our original problem. The variational parameters to be
adjusted are $\lambda = \{ \mu_{1\:H}, \Sigma_{1\:H} \}$. We conduct the
variational inference~\myeqref{elboopt1} using block coordinate ascent on
the coordinate blocks $\mu_1$, $\Sigma_1$, \dots, $\mu_H$, $\Sigma_H$. The
coordinate updates do not have a closed form, but each update solves a
smooth, unconstrained convex optimization problem, as we show
in~\myapx{varinf}. There we also give a closed-form gradient and Hessian
for the $\mu_h$ update, as well as closed-form gradients for the $\Sigma_h$
update under two different parametrizations.

To finish with empirical Bayes, we explain how to obtain approximate MLEs
$\hat{\zeta}$ and $\hat{\Omega}$. Notice the variational inference
procedure~\myeqref{elboopt1} yields an optimal value $\lambda^*$ for fixed
$\zeta$ and $\Omega$. We alternate this variational inference step with a
complementary optimization over $\zeta$ and $\Omega$. In fact, these
optimizations constitute a version of the expectation-maximization (EM)
algorithm for computing MLEs. The standard E-step, where we compute the
posterior expected complete log likelihood, is replaced with a variational
E-step, where the expected complete log likelihood is approximated using
$\qr{ \beta_{1\:H} \g \lambda^*}$. The variational EM algorithm alternates
between the following steps until $\hat{\zeta}$, $\hat{\Omega}$, and the
variational parameters stabilize:
\begin{description}

\item[E-step (variational).] Using the current $\hat{\zeta}$ and
  $\hat{\Omega}$, run the block coordinate ascent algorithm as described
  in~\myapx{varinf}, yielding new variational parameter values $\{
  \mu_{1\:H}^*, \Sigma_{1\:H}^* \}$.

\item[M-step.] Using the current variational parameter values, update the
  empirical Bayes parameter estimates: $(\hat{\zeta}, \hat{\Omega})
  \leftarrow \argmax_{\zeta, \Omega} \ \mathcal{L}(\mu_{1\:H}^*,
  \Sigma_{1\:H}^* ; \zeta, \Omega)$.

\end{description}
It can be shown that the variational E-step finds a new value of $\lambda$
which moves the lower-bounding function $\mathcal{L}(\lambda; \zeta,
\Omega)$ towards the true log-likelihood $\log p(\D \g \zeta, \Omega)$ from
below. The M-step maximizes this adjusted lower-bounding function over
$\zeta$ and $\Omega$, as a surrogate for the true log-likelihood. We then
re-tighten and re-maximize until convergence. \myapx{varinf} gives details
on initialization and the M-step update.

\subsection{Variational hierarchical Bayes}

Fully Bayesian inference in the mixed multinomial logit model requires
calculations under the posterior $\pr{ \beta_{1\:H} , \zeta , \Omega \g \D
}$ given in~\myeqref{hbpost}. In one sense, the setting is simpler than
empirical Bayes: there are no unknown top-level parameters to estimate. All
we need is to extend the previous section's variational inference procedure
to include $\zeta$ and $\Omega$. \myapx{varinf-hb} reports the details
behind the extension; here we summarize the main ideas.

Although the joint posterior~\myeqref{hbpost} is not factorized, we
continue to use a family $\mathcal{Q}$ of factorized distributions for the
variational approximation:
\begin{equation}
  \mathcal{Q} \ni \qr{\beta_{1\:H}, \zeta, \Omega \g \lambda} \defsym
  \qr{\zeta \g \mu_\zeta, \Sigma_\zeta} \qr{\Omega \g \Upsilon^{-1}, \omega}
  \prod_{h=1}^H \qr{\beta_h \g \mu_h, \Sigma_h} . \label{eq:hb-q}
\end{equation}
Using a factored family for a non-factored posterior is commonly called
\textit{mean-field} variational inference. In~\myeqref{hb-q}, $\qr{\zeta \g
  \mu_\zeta, \Sigma_\zeta}$ is a $K$-variate normal density; $\qr{\Omega \g
  \Upsilon^{-1}, \omega}$ is an inverse Wishart density; and the
$q(\beta_h)$ factors are $K$-variate normal densities as before. In the
analysis and the algorithm, it is convenient to use a well-known
equivalence, treating $\qr{\Omega \g \Upsilon^{-1}, \omega}$ as a Wishart
distribution $\mathcal{W}(\Upsilon, \omega)$ on $\Omega^{-1}$. We are
therefore optimizing over variational parameters $\lambda \defsym \left\{
  \mu_\zeta, \Sigma_\zeta, \Upsilon, \omega, \mu_{1\:H}, \Sigma_{1\:H}
\right\}$. The variational problem for hierarchical Bayes is to find the
best approximating distribution $\qr{\beta_{1\:H}, \zeta, \Omega \g
  \lambda^*}$ in $\mathcal{Q}$ by solving the analog of~\myeqref{elboopt1}:
\begin{equation}
  \lambda^* = \argmax_{\lambda \in \Lambda} \ \mathcal{L}(\lambda) \ .
  \label{eq:elboopt-hb}
\end{equation}
As with empirical Bayes, we use a block coordinate ascent optimization
algorithm to solve~\myeqref{elboopt-hb}, iterating through the coordinate
blocks that define $\lambda$. Here again, all coordinate updates are
convex optimizations. The details appear in~\myapx{varinf}: updates for
$\mu_\zeta$, $\Sigma_\zeta$, $\Upsilon$ all have simple closed forms;
$\omega$ has a closed form which requires no updating; and the $\mu_h$ and
$\Sigma_h$ updates are similar to the empirical Bayes case.

\section{Empirical results} \label{sec:results}

We compared the accuracy and speed of the variational methods described in
the previous section to a standard and widely used MCMC
approach~\citep{AllenbyRossi2003}, on a suite of simulated data sets. Each
data set was generated using the discrete choice model given
by~\myeqref{yht} and~\myeqref{betah}. To simulate a data set with $J$
choice items, $K$ item attributes, and $H$ agents, we first fixed values of
$\zeta$ and $\Omega$, the parameters controlling the distribution of
preferences in the agent population. We then independently drew a $\beta_h$
vector for each agent, according to~\myeqref{betah}. We also drew for each
agent 25 iid $J \times K$ item attribute matrices $x_{ht}$ consisting of
iid $N(0,0.5^2)$ entries. Finally, for each agent, we used $x_{ht}$ and
$\beta_h$ to simulate 25 choice events $y_{ht}$, according
to~\myeqref{yht}. Thus, in our data sets, each agent has $25$ observed
choices.

We simulated a total of 32 different scenarios by varying $J$, $K$, $H$,
and the selection of $\zeta$ and $\Omega$. Specifically, each data set
corresponds to a distinct configuration of the following candidate values:
3 or 12 choice items $J$; 3 or 10 item attributes $K$; 250, 1000, 5000, or
25000 agents $H$; and ``low'' or ``high'' heterogeneity of the agent
population. In the low-heterogeneity scenario, the $K \times 1$ vector
$\zeta$ consists of evenly spaced values from -2 to 2, and the $K \times K$
matrix $\Omega$ is 0.25 times the identity matrix. In the
high-heterogeneity scenario, $\zeta$ is the same, but $\Omega$ is the
identity matrix. The data sets with high heterogeneity have much more
diverse collections of preference vectors $\beta_h$.

We ran variational empirical Bayes (VEB), variational hierarchical Bayes
(VB), and the standard MCMC algorithm on the observable data from each of
the 32 simulation scenarios. For VEB, we declared convergence as soon as an
E-step/M-step iteration caused the parameter estimates' joint Euclidean
norm to change by less than $10^{-4}$, relative to their norm at the
beginning of the iteration (here we mean the joint norm of all the
variational parameters, together with the model parameter estimates
$\hat{\zeta}$ and $\hat{\Omega}$). The convergence criterion for VB was the
same, except $\zeta$ and $\Omega$ do not have point estimates -- we look
instead at the change in the variational parameters corresponding to their
posterior approximation, namely $\mu_\zeta$, $\Sigma_\zeta$, and
$\Upsilon$.

Choosing MCMC convergence criteria is more delicate. We tried to set the
number of burn-in iterations and the thinning ratio algorithmically, using
the technique of~\citet{RafteryLewis1992}. But on several of our data sets,
typical control parameter values, such as the default settings for the
\texttt{raftery.diag} function in the R package
\texttt{coda}~\citep{R_coda}, led to a very large number of burn-in
iterations. Trace plots of the sampled parameters indicated these large
burn-in values were unnecessary, so using them would have been unfair to
MCMC in our timing comparisons. Instead, we manually investigated MCMC
convergence and autocorrelation for several data sets, using trace plots,
partial autocorrelation functions, and related diagnostics available in the
\texttt{coda} package. Based on these studies, we chose to use 1,000
iterations of burn-in and a 10:1 thinning ratio. On each data set, we
therefore ran 6,000 total iterations of MCMC to generate 500 draws for the
approximate posterior. These numbers are as small as we could reasonably
make them, in order to be fair to MCMC in the timing comparisons.

\subsection{Accuracy}

Our measure of accuracy for each inference procedure is based on the
\textit{predictive choice distribution}. Informally, this distribution
gives the item choice probabilities exhibited by the ``average'' agent,
when shown an item attribute matrix $x_{\text{new}}$. In our simulations,
we know the true values of $\zeta$ and $\Omega$, so we can compute the true
predictive choice distribution:
\begin{align}
  \pr{ y_{\text{new}} \g x_{\text{new}} , \zeta, \Omega } & =
  \int \pr{ y_{\text{new}} \g x_{\text{new}} , \beta } \pr{\beta \g
    \zeta, \Omega } d \beta \label{eq:true-predictive-choiceprobs1} \\
  & = \E_\beta \pr{ y_{\text{new}} \g x_{\text{new}} , \beta} \ .
  \label{eq:true-predictive-choiceprobs2}
\end{align}
Equation~\myeqref{true-predictive-choiceprobs2} explains the ``average
agent'' interpretation of the predictive choice distribution. A slightly
different take
on~\myeqref{true-predictive-choiceprobs1}-\myeqref{true-predictive-choiceprobs2}
is the following: if we want to forecast the item choice probabilities of a
new, previously unobserved decision-maker, we can think of her as a random
draw from the agent population. Under our model, the choice probabilities
for such a randomly drawn agent are
precisely~\myeqref{true-predictive-choiceprobs1}-\myeqref{true-predictive-choiceprobs2}.

For any particular $x_{\text{new}}$, we use the true values of $\zeta$ and
$\Omega$ to compute a Monte Carlo estimate
of~\myeqref{true-predictive-choiceprobs2}. We base the estimate on enough
draws of $\beta \sim \mathcal{N}(\zeta, \Omega)$ to insure that its
variability does not affect even the least significant digit of our
reported results. We handle the integral over $\beta$ in the same way for
the estimated predictive choice distributions furnished by each of the
three inference procedures. For VEB, the estimate
is~\myeqref{true-predictive-choiceprobs1}, with $\zeta$ and $\Omega$
replaced by $\hat{\zeta}_{\text{VEB}}$ and $\hat{\Omega}_{\text{VEB}}$. On
the other hand, with VB and MCMC, we obtain a posterior distribution over
$\zeta$ and $\Omega$; we take the mean
of~\myeqref{true-predictive-choiceprobs1} under this posterior as a point
estimate of the predictive choice distribution:
\begin{equation}
  \hat{p} \!\left( y_{\text{new}} \g x_{\text{new}}, \D \right) =
  \int \left[ \int \pr{ y_{\text{new}} \g x_{\text{new}} , \beta } \pr{\beta \g
      \zeta, \Omega } d \beta \right] \pr{\zeta, \Omega \g \D}
  d\zeta \, d\Omega \ . \label{eq:posterior-predchoice}
\end{equation}
For VB, the posterior density $\pr{\zeta, \Omega \g \D}$
in~\myeqref{posterior-predchoice} is approximated by the fitted variational
distribution
\begin{equation}
\qr{\zeta \g \mu_\zeta, \Sigma_\zeta} \qr{\Omega \g \Upsilon^{-1}, \omega} \ .
\end{equation}
For MCMC, the posterior is approximated as usual by the empirical
distribution of draws from a Markov chain. In both cases, we handle the
integral over $\zeta$ and $\Omega$ in~\myeqref{posterior-predchoice} with
another exhaustive Monte Carlo approximation.

We measure the error of each inference procedure as the distance from its
estimate of the predictive choice distribution to the true distribution. As
the metric on distributions, we use total variation (TV) distance, leading
to what we call the ``TV error'' of each procedure. In this setting, TV
error equals the maximum, over all choice-item subsets, of the difference
in the probabilities assigned to the subset by the estimated versus the
true choice distribution. We also need to choose the attribute matrix
$x_{\text{new}}$ at which the true and estimated predictive choice
distributions are calculated. In each simulation scenario, we computed the
TV error of VEB on 25 different random draws of $x_{\text{new}}$. We then
compared the three procedures using the $x_{\text{new}}$ which yielded the
median of the 25 TV errors. In this sense our results are representative of
accuracy for a ``typical'' item attribute matrix. However, results using
any one of the 25 matrices were qualitatively the same in the examples we
checked.

TV error values for the three inference procedures are presented in
Table~\mytbl{3item} for the 3-item simulation scenarios, and in
Table~\mytbl{12item} for the 12-item scenarios. The main conclusion we draw
from Tables~\mytbl{3item} and~\mytbl{12item} is simple: on these data sets,
there are no practical differences in accuracy among VEB, VB, and MCMC. The
scale of the TV error for all the procedures is the same; that scale is
larger in the 12-item case than the 3-item case, but all three procedures
exhibit high accuracy on all data sets. The magnitude of our simulation
study makes it difficult to carry out the replications required to put
standard errors in these tables. But even if the differences among TV
errors in every scenario were ``significant'' under a suitable definition,
the patternless alternation in the identity of the most accurate method
would make more specific conclusions dubious.

\begin{table}
  \begin{center}
    \begin{tabular}{rl*{7}{r}}
      && \multicolumn{3}{c}{3 attributes} & \quad & \multicolumn{3}{c}{10 attributes} \\ \cline{3-5} \cline{7-9}
      \rule{0ex}{1em} Agents & Het. & VEB & \quad\ \ VB & MCMC & \quad & VEB & \quad\ \ VB & MCMC \\ \hline
      \multirow{2}{*}{250} & Low \rule{0ex}{1em}  & 0.23 & 0.31 & 0.35 & \quad & 0.66 & 0.50 & 0.59 \\
      & High & 0.49  & 0.48 & 0.92 & \quad & 0.68 & 0.51 & 1.46 \\ \hline
      \multirow{2}{*}{1,000} & Low \rule{0ex}{1em} & 0.74 & 0.63 & 0.46 & \quad & 0.49 & 0.17 & 0.14 \\
      & High & 0.53 & 0.33 & 0.09 & \quad & 0.30 & 0.39 & 0.18 \\ \hline
      \multirow{2}{*}{5,000} & Low \rule{0ex}{1em} & 0.25 & 0.37 & 0.35 & \quad & 0.56 & 0.59 & 0.07 \\
      & High & 0.19 & 0.25 & 0.39 & \quad & 0.35 & 0.20 & 0.38 \\ \hline
      \multirow{2}{*}{25,000} & Low \rule{0ex}{1em} & 0.63 & 0.74 & NA & \quad & 0.59 & 0.55 & NA \\
      & High & 0.53 & 0.53 & NA & \quad & 1.60 & 1.10 & NA
    \end{tabular}
    \caption{Total variation error in percentage points, for simulated data
      sets with three choice items. MCMC results are unavailable in the
      25,000 agent case because the sampler exhausted memory resources
      before converging. See the text for the definition of total variation
      error.}
    \label{tbl:3item}
  \end{center}
\end{table}

\begin{table}
  \begin{center}
    \begin{tabular}{rl*{7}{r}}
      && \multicolumn{3}{c}{3 attributes} & \quad & \multicolumn{3}{c}{10 attributes} \\ \cline{3-5} \cline{7-9}
      \rule{0ex}{1em} Agents & Het. & VEB & \quad\ \ VB & MCMC & \quad & VEB & \quad\ \ VB & MCMC \\ \hline
      \multirow{2}{*}{250} & Low \rule{0ex}{1em}  & 2.88 & 2.80 & 2.64 & \quad & 1.97 & 1.91 & 2.44 \\
      & High & 1.44  & 1.94 & 1.64 & \quad & 2.43 & 2.37 & 2.62 \\ \hline
      \multirow{2}{*}{1,000} & Low \rule{0ex}{1em} & 1.11 & 1.27 & 1.09 & \quad & 0.99  & 1.00 & 1.60 \\
      & High & 0.98 & 1.18 & 1.18 & \quad & 1.99 & 2.05 & 1.96 \\ \hline
      \multirow{2}{*}{5,000} & Low \rule{0ex}{1em} & 1.25 & 1.45 & 1.18 & \quad & 0.95 & 1.13 & 0.97 \\
      & High & 1.14 & 0.98 & 1.10 & \quad & 0.71 & 0.91 & 0.92 \\ \hline
      \multirow{2}{*}{25,000} & Low \rule{0ex}{1em} & 0.22 & 0.33 & NA & \quad & 0.51 & 0.53 & NA \\
      & High & 0.99 & 0.57 & NA & \quad & 1.23 & 0.96 & NA
    \end{tabular}
    \caption{Total variation error in percentage points, for simulated data
      sets with 12 choice items. See the caption accompanying
      Table~\mytbl{3item}.}
    \label{tbl:12item}
  \end{center}
\end{table}

\begin{figure}
  \begin{center}
    \includegraphics[scale=.6]{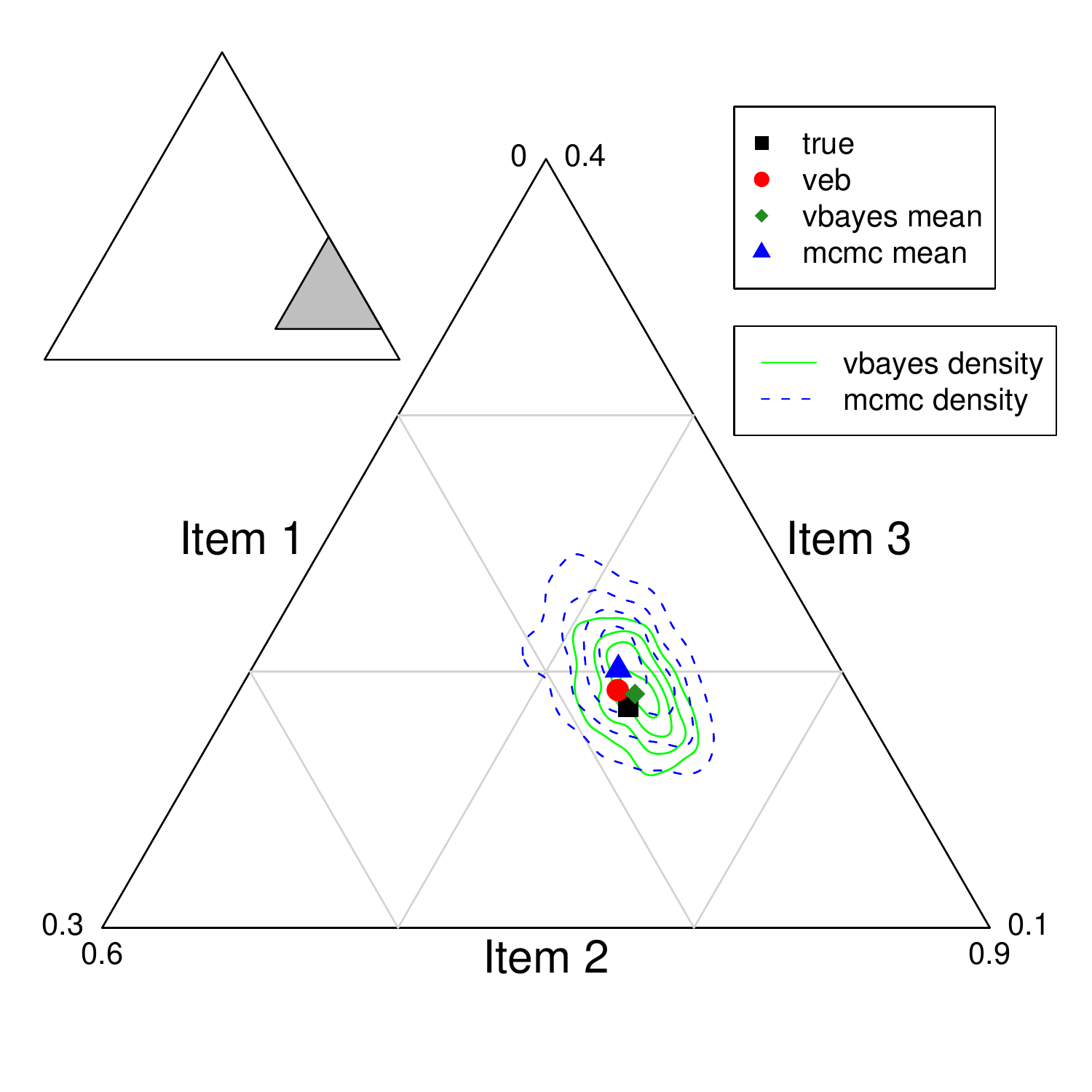}
    \caption{Triangle plot of the true predictive choice distribution and
      its estimates in the three-item case, for the simulation with 250
      decision-makers, 10 attributes, and high heterogeneity. See
      accompanying text.}
    \label{fig:A1003H}
  \end{center}
\end{figure}

Figure~\myfig{A1003H} shows in a different way the comparable accuracy of
the three procedures. When there are three choice items, any predictive
choice distribution can be plotted as a point in the triangular simplex.
The figure shows the close proximity of each procedure's estimated choice
distribution to the true distribution in one simulation scenario. Plots for
all the other three-item scenarios are qualitatively the same.
Figure~\myfig{A1003H} also shows contours of the VB and MCMC approximate
posterior distributions. We see that VB is producing not only a posterior
mean similar to MCMC, but also a similar posterior density in the
neighborhood of the mean.

\subsection{Speed}

For each simulated data set, we ran the three procedures in turn on the
same unloaded machine: a 64-bit dual-core 3.2 GHz Intel Xeon processor with
8 GB of main memory. For the MCMC inference, we used the
\texttt{rhierMnlRwMixture} function in the R package
\texttt{bayesm}~\citep{R_bayesm}, which has efficient vectorized
implementations of the inner sampling routines. This package stores all
MCMC draws in memory, however. For our largest data sets, with 25,000
decision-makers, the machine's memory was exhausted before MCMC converged.
We were able to run MCMC for 1,000 iterations in this case, which allowed
us to extrapolate accurately the time that would have been required for
6,000 iterations. We implemented the variational algorithms in R, with
compiled C components for the numerical optimization routines.

\begin{figure}
  \begin{center}
    \includegraphics[scale=0.85]{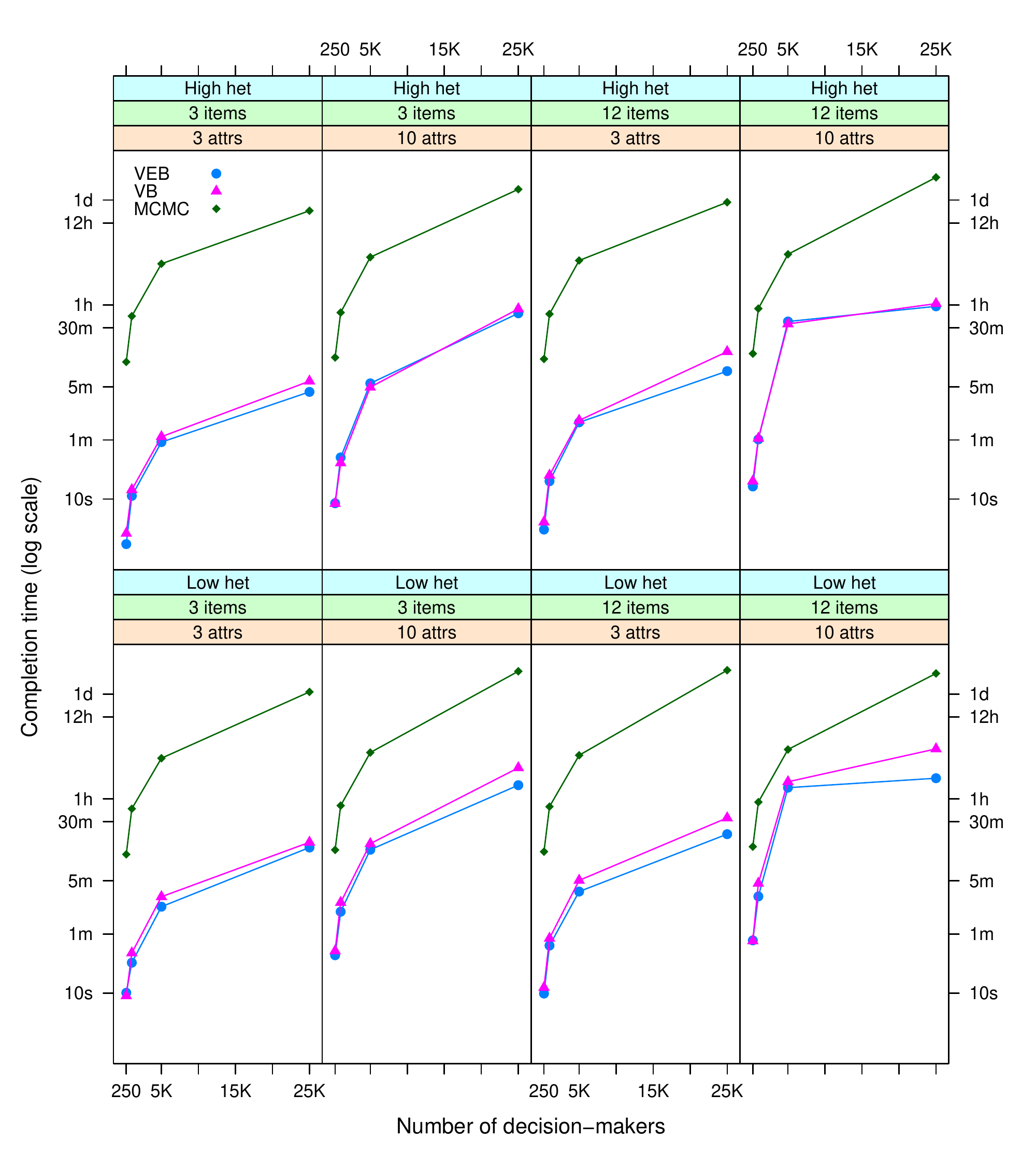}
    \caption{Timing results for variational empirical Bayes (VEB),
      variational hierarchical Bayes (VB), and MCMC. Within each panel,
      convergence time is plotted on the log scale as a function of the
      number of agents, for fixed values of the other simulation parameters
      (shown at the top of each panel).}
    \label{fig:timing}
  \end{center}
\end{figure}

Figure~\myfig{timing} displays time to convergence on each data set for the
three procedures, according to the convergence criteria previously
described. Within each panel, convergence time is plotted as a function of
the number of agents, for fixed values of the other simulation parameters.
Note that the vertical axis shows convergence time on a logarithmic scale,
to ease comparison of MCMC to the variational methods. All the procedures
scale roughly linearly with the number of agents, which leads to the
logarithmic curves seen in the figure. The conclusions are the same in all
the scenarios we simulated: variational methods converge faster than MCMC,
and the magnitude of the difference increases with the number of observed
agents. MCMC uses two days of computation time for 25,000 agents with 12
choice items, 10 item attributes, and high heterogeneity, versus an hour
for each of the variational techniques. In the same setting but with low
heterogeneity, MCMC's two-day computation compares with two hours for VEB
and six hours for VB. In other scenarios, variational run times are
measured in minutes, as opposed to hours or days for MCMC.

\section{Discussion}

\label{sec:discussion}

Variational methods allow estimation of hierarchical discrete choice models
in a small fraction of the time required for MCMC. They open Bayesian
inference to a wide class of problems for which resource constraints make
MCMC intractable, such as the MML model with many heterogeneous units. For
now, variational methods appear to be the only viable option in these
cases.

Of course, one can use variational methods to estimates many more types of
models than the MML model examined here. Within the MML family, it would be
straightforward to add a utility scaling parameter, or allow the
heterogeneous coefficients themselves to depend on observed covariates. The
value of the variational approach is greatest when subsampling of data is
ill-advised. For example, consider a linear model with heterogeneous
coefficients on exogenous and endogenous covariates, where the available
instrumental variables only weakly explain the endogenous part. To draw
inferences about the covariances of the heterogeneous parameters, we may
need a large amount of data to achieve reasonable power in hypothesis
testing. MCMC is untenable here, but variational methods have promise. When
factorized variational distributions are inadequate, alternatives such as
mixtures of normals or Dirichlet processes~\citep{BleiJordan2006} can be
applied.

We emphasize that we do not advocate abandoning MCMC in favor of
variational methods. On the contrary, we suggest using MCMC when possible.
MCMC offers consistency guarantees with no analog in variational methods,
and it has withstood decades of scrutiny. But we advise against subsampling
the data (i.e., throwing out information), or discarding key modeling
elements, simply to make the problem fit within the time and resource
constraints of MCMC. The possibility of applying variational methods to
previously intractable problems makes them an important addition to the
statistician's toolkit.

\appendix

\section{Variational inference and parameter estimation}
\label{apx:varinf}

In this appendix we describe the variational inference and estimation
procedures for the mixed multinomial logit model. A few words on notation:
$A \succ 0$ means the matrix A is positive definite; $A \succeq 0$ means
positive semidefinite; $|A|$ is the determinant of $A$. For a scalar
function $s \ \:\ \mathbb{R} \to \mathbb{R}$ and a vector $v \in
\mathbb{R}^n$, $s(v)$ means the $n \times 1$ vector $(s(v_1), \ldots,
s(v_n))^\top$.

\subsection{The empirical Bayes ELBO}

For empirical Bayes in the MML model, the ELBO objective
function~\myeqref{elbo} becomes
\begin{equation}
  H(q) +
  \sum_{h=1}^H \E_q \log \pr{\beta_h \g \zeta, \Omega} +
  \sum_{h=1}^H \sum_{t=1}^{T_h} \E_q \log \pr{y_{ht} \g x_{ht},
    \beta_h} \ . \label{eq:elbo3}
\end{equation}
The first term in~\myeqref{elbo3} is the Shannon entropy of the variational
distribution. The second and third terms are (minus) an unnormalized cross
entropy -- the missing normalization constant is the marginal likelihood.
Recall that the variational distribution $\qr{\beta_{1\:H} \g \mu_{1\:H},
  \Sigma_{1\:H}}$ is a product of normal distributions $\mathcal{N}(\mu_h,
\Sigma_h)$.

The first and second terms of~\myeqref{elbo3} are straightforward to
derive; the third term requires more attention. Using the multinomial logit
mass function
\begin{equation}
  \pr{y_{ht} \g x_{ht}, \beta_h} = \prod_{j=1}^J \left[
    \frac{ \exp(x_{htj}^\top \beta_h) }{ \sum_{j'} \exp(x_{htj'}^\top \beta_h) }
  \right]^{y_{ht}^j} \ ,
\end{equation}
the third term becomes
\begin{equation}
  \sum_{h=1}^H \sum_{t=1}^{T_h}
  \left[
    \sum_{j=1}^J y_{ht}^j (x_{htj}^\top \mu_h) -
    \E_{q_\lambda} \log\biggl(\sum_{j=1}^J \exp(x_{htj}^\top \beta_h) \biggr)
  \right] \ . \label{eq:lse}
\end{equation}
The expected log-sum-exp in~\myeqref{lse} has no closed form. For
variational inference, we therefore approximate the ELBO objective function
$\mathcal{L}$ using a new objective function $\tmL$. To construct $\tmL$,
we consider two alternatives: the zeroth-order and first-order delta method
for moments~\citep{BickelDoksum2007}, which we call D0 and D1 respectively.
D0 is equivalent to applying Jensen's inequality to the expected
log-sum-exp, resulting in the following lower bound to~\myeqref{lse}:
\begin{equation}
  \textbf{[D0]} \quad
  \sum_{h=1}^H \sum_{t=1}^{T_h}
  \left[
    \sum_{j=1}^J y_{ht}^j (x_{htj}^\top \mu_h) -
    \log \biggl(
    \sum_{j=1}^J \exp( x_{htj}^\top \mu_h + (1/2) x_{htj}^\top \Sigma_h x_{htj} )
    \biggr)
  \right] \ . \label{eq:lseD0}
\end{equation}
Here we used the usual formula for the mean of a lognormal random variable.
In a different context, \citet{BleiLafferty2007} consider an approximation
equivalent to D0 but expressed using a redundant variational parameter.

For approximation D1, we restrict $\Sigma_h \succ 0$ to be diagonal, and
define \\
$\sigma_h \defsym \log(\diag\{ \Sigma_h \}) \in \mathbb{R}^K$. Using
results in \myapx{delta-method}, we obtain the following approximation
to~\myeqref{lse}:
\begin{flalign}
  \textbf{[D1]} \quad
  \sum_{h=1}^H \sum_{t=1}^{T_h}
  \left[
    \sum_{j=1}^J y_{ht}^j (x_{htj}^\top \mu_h) -
    \log \biggl( \sum_{j=1}^J \exp( x_{htj}^\top \mu_h ) \biggr)
    - \frac{1}{2} \exp(\sigma_h)^\top \Theta(\mu_h)
  \right] \ , \label{eq:lseD1}
\end{flalign}
with $\Theta(\mu_h) \in \mathbb{R}^K$ as defined in \myapx{delta-method}.
Notice that, unlike D0, approximation D1 does not preserve the guarantee
that the optimal value of the variational optimization lower bounds the
marginal likelihood. However, in our simulations, using D1 resulted in more
accurate variational approximations to the posterior.

In this appendix we give a derivation based on approximation D0. The
derivation for D1 is similar, but simpler, because $\Sigma_h$ is treated as
diagonal. Under D0, the final empirical Bayes objective function is
\begin{equation} \label{eq:tmLD0}
  \begin{split}
    & \tmL(\mu_{1\:H}, \! \Sigma_{1\:H} ; \, \zeta, \Omega) =
    \frac{1}{2} \sum_{h=1}^H \log\left[(2\pi e)^K |\Sigma_h|\right] \\
    & \quad - \frac{H}{2} \log \left( (2\pi)^K |\Omega| \right)
    - \frac{1}{2} \tr
    \left[
      \Omega^{-1}
      \sum_{h=1}^H \left\{
        \Sigma_h + (\mu_h - \zeta) (\mu_h - \zeta)^\top
      \right\}
    \right] \\
    & \quad + \sum_{h=1}^H \sum_{t=1}^{T_h}
    \left[
      \sum_{j=1}^J y_{ht}^j (x_{htj}^\top \mu_h) -
      \log \biggl(
      \sum_{j=1}^J \exp( x_{htj}^\top \mu_h +
      (1/2) x_{htj}^\top \Sigma_h x_{htj} )
      \biggr)
    \right] \ .
  \end{split}
\end{equation}
The first line in~\myeqref{tmLD0} uses the well-known entropy of the normal
distribution. The second line uses the cross-entropy of two normal
distributions, also well known. The third line is approximation D0.

\subsection{Empirical Bayes variational E-step}

Here we describe a block coordinate ascent algorithm to
maximize~\myeqref{tmLD0} over the variational parameters $\mu_{1\:H}$ and
$\Sigma_{1\:H}$. Although the problem is not jointly convex in all these
parameters, each $\mu_h$ and $\Sigma_h$ coordinate update solves a smooth,
unconstrained convex optimization problem. The requirement $\Sigma_h
\succeq 0$ is satisfied after each update. We initialize the variational
parameters at the maximum likelihood estimates from a homogeneous model (in
which all agents share a common $\beta$ value).

The concavity of~\myeqref{tmLD0} in $\mu_h$ follows from the fact that
$\Omega \succ 0$ and from the convexity of the log-sum-exp function. We
update $\mu_h$ using standard algorithms for unconstrained convex
optimization~\citep{BoydVandenberghe2004}, supplying an analytic gradient
and Hessian as follows. Define the function $w(\mu,\Sigma,x)$ taking values
in $\mathbb{R}^J$, with $j$th component $\exp \left( x_j^\top \mu + (1/2)
  x_j^\top \Sigma x_j \right)$, and normalized to sum to one across $j$.
The gradient of $\tmL$ with respect to $\mu_h$ can then be written
\begin{equation}
  \frac{\partial \tmL}{\partial \mu_h} =
  - \Omega^{-1} (\mu_h - \zeta) +
  \sum_{t=1}^{T_h} \sum_{j=1}^J
  \left[ y_{ht}^j  - w^j(\mu_h, \Sigma_h, x_{ht}) \right]
  x_{htj} \ . \label{eq:eb-muh-gradient}
\end{equation}
Note the similarity of this gradient to the gradient from an
$L_2$-regularized multiple logistic regression: it consists of a
contribution from the regularizer (the left-hand term), plus a
residual-weighted sum of covariate vectors. Abbreviating $w(\mu_h,
\Sigma_h, x_{ht})$ to $w_{ht}$, an argument using matrix
differentials~\citep{MagnusNeudecker2007} gives the Hessian
\begin{equation}
  \frac{\partial \tmL}{\partial \mu_h \partial \mu_h^\top} =
  - \Omega^{-1}
  - \sum_{t=1}^{T_h} \left[
    x_{ht}^\top \, \text{diag}\{w_{ht}\} \, x_{ht}
    - (x_{ht}^\top w_{ht}) (x_{ht}^\top w_{ht})^\top
    \right] \ . \label{eq:eb-muh-hessian}
\end{equation}

The $\Sigma_h$ coordinate update is harder, because we need to insure that
$\Sigma_h \succeq 0$. Using a reformulation, we can avoid making the
constraint explicit, which would complicate the optimization. Let $\Sigma_h
= L_h L_h^\top$ for a lower-triangular matrix $L_h$. Since $\Sigma_h
\succeq 0$, one such $L_h$ always exists---the Cholesky factor. We replace
each $\Sigma_h$ in $\tmL$ with $\Sigma_h(L_h) \defsym L_h L_h^\top$, and
optimize over the unconstrained set of lower-triangular matrices $L_h$.

The objective function~\myeqref{tmLD0} remains concave in $L_h$. To see
this, compare the terms depending on $\Sigma_h = L_h L_h^\top$ to the
function studied in~\myapx{convexity}. We now give the gradient with
respect to $L_h$. Standard matrix differentiation of~\myeqref{tmLD0} leads
to the $\Sigma_h$ gradient
\begin{equation}
  \frac{\partial \tmL}{\partial \Sigma_h} =
  \frac{1}{2} \left[
    \Sigma_h^{-1} - \Omega^{-1}
    - \sum_{t=1}^{T_h} x_{ht}^\top \text{diag}\{w_{ht}\} x_{ht}
  \right] \ . \label{eq:sigmagrad}
\end{equation}
Again using matrix differentials and the Cauchy invariance rule, it is not
hard to show that the gradient with respect to $L_h$ is
\begin{equation}
  \frac{\partial \tmL}{\partial L_h}
  = 2 \left( \frac{\partial \tmL}{\partial \Sigma_h} \right) L_h
  = L_h^{-\top} - \left( \Omega^{-1} +
    \sum_{t=1}^{T_h} x_{ht}^\top \text{diag}\{w_{ht}\} x_{ht} \right) L_h \ .
  \label{eq:eb-lh-gradient}
\end{equation}
Note that this is the gradient with respect to a dense matrix $L_h$. Since
we optimize over lower-triangular matrices, i.e. vech($L_h$), we need only
use the lower triangular of the gradient. This is convenient for
the term $L_h^{-\top}$: it is upper-triangular, so its lower triangle is
a diagonal matrix. Furthermore, from a standard result of linear algebra,
the diagonal entries are simply $1/\ell_{ii}$, where the $\ell_{ii}$'s form
the diagonal of $L_h$.

In practice we do the $\mu_h$ and $\Sigma_h$ updates in a single step by
optimizing jointly over $\mu_h$ and $L_h$, which remains a convex problem.

\subsection{Empirical Bayes M-step}

In the M-step, we maximize~\myeqref{tmLD0} over $\zeta$ and $\Omega$.
Identifying the terms which depend on $\zeta$, we recognize the usual
Gaussian mean estimation problem. Further,~\myeqref{tmLD0} is easily seen
to be concave in $\Omega^{-1}$, with a closed-form solution of the
corresponding first-order condition. We obtain the M-step updates
\begin{align}
  \hat{\zeta} & \leftarrow \frac{1}{H} \sum_{h=1}^H \mu_h \ , &
  \hat{\Omega} & \leftarrow \frac{1}{H} \sum_{h=1}^H \Sigma_h
  + \widehat{\text{Cov}}(\mu_{\cdot}) \ .
\end{align}
Here $\widehat{\text{Cov}}(\mu_{\cdot})$ is the empirical covariance of the
$\mu_h$ vectors.

\subsection{Variational hierarchical Bayes} \label{apx:varinf-hb}

In the fully Bayesian MML model, $\zeta$ and $\Omega$ have prior
distributions, with corresponding variational factors given
in~\myeqref{hb-q}. The ELBO in this case has the same form
as~\myeqref{elbo3}, with two differences. First, $H(q)$ contains two new
terms
\begin{equation}
  H\left(\qr{\zeta \g \mu_\zeta, \Sigma_\zeta}\right) +
  H\bigl(q\bigl(\Omega \g \Upsilon^{-1}, \omega\bigr)\bigr) \ .
\end{equation}
Second, there are two new cross-entropy terms
\begin{equation}
  \E_q \log \pr{\zeta \g \beta_0, \Omega_0} +
  \E_q \log \pr{\Omega \g S, \nu} \ .
\end{equation}
Also, the middle term of~\myeqref{elbo3} changes in the fully Bayesian
case, because $\zeta$ and $\Omega$ are now averaged over rather than
treated as constants.

Using known formulas for normal and Wishart entropies, the two new entropy
terms are seen to equal
\begin{equation}
  \frac{1}{2}\log\left[(2\pi e)^K |\Sigma_\zeta|\right]
  - \frac{\omega-K-1}{2} D(\omega,\Upsilon) + \frac{\omega K}{2} +
  A_\omega\bigl(\Upsilon\bigr) \ .
\end{equation}
Here we used the expected log determinant of a Wishart random matrix
\begin{equation}
  D(\omega,\Upsilon) \defsym \log \left( 2^K |\Upsilon| \right) + \sum_{i=1}^K
  \Psi \left( \frac{\omega+1-i}{2} \right)
\end{equation}
and the log normalization constant of the Wishart distribution
\begin{equation}
  A_\omega(\Upsilon) \defsym \log \left[ 2^{\omega K / 2} \pi^{K(K-1)/4}
    \prod_{i=1}^K \Gamma\left(\frac{\omega+1-i}{2}\right) \right] +
  \frac{\omega}{2} \log |\Upsilon|
\end{equation}
\citep[see, for example,][]{Beal2003}. The new cross entropy terms for
$\zeta$ and $\Omega$ work out to
\begin{equation}
-\frac{1}{2} \left\{ \log\left[(2\pi)^K |\Omega_0|\right]
    + \tr\left( \Omega_0^{-1} \left[ \Sigma_\zeta +
        (\mu_\zeta - \beta_0)(\mu_\zeta - \beta_0)^\top \right] \right) \right\}
\end{equation}
and
\begin{equation}
  -A_\nu\bigl(S^{-1}\bigr) + \frac{\nu-K-1}{2} D(\omega, \Upsilon) -
  \frac{\omega}{2} \tr\bigl(S^{-1} \Upsilon\bigr)
\end{equation}
respectively. The middle term of~\myeqref{elbo3} eventually becomes
\begin{multline}
  - \frac{H}{2} \left\{ K \log(2\pi) - D(\omega, \Upsilon) \right\} \ - \\
  \frac{\omega}{2}
  \tr \left[
    \Upsilon
    \left(
      H \Sigma_\zeta +
      \sum_{h=1}^H
      \left( \Sigma_h + (\mu_\zeta - \mu_h)(\mu_\zeta - \mu_h)^\top \right)
    \right)
  \right] \ .
\end{multline}

With these changes, it is not hard to see that $\tmL$ is concave separately
in $\mu_\zeta$ and $\Sigma_\zeta$. The first-order conditions for block
coordinate ascent lead to the updates
\begin{align}
  \mu_\zeta & \leftarrow\ \bigl(\Omega_0^{-1} + H \omega
  \Upsilon \bigr)^{-1} \biggl(\Omega_0^{-1} \beta_0 +
  \omega \Upsilon \sum_{h=1}^H \mu_h\biggr) \ , \\
  \Sigma_\zeta & \leftarrow\ \bigl(\Omega_0^{-1} + H \omega
  \Upsilon \bigr)^{-1} \ .
\end{align}
By inspection, $\Sigma_\zeta \succeq 0$, so this constraint need not be
explicitly enforced. Note the similarity to conjugate posterior updating:
on the precision scale, $\Sigma_\zeta$ is the sum of the prior precision
matrix $\Omega_0^{-1}$ and $H$ copies of the variational posterior mean
$\omega \Upsilon$ for $\Omega^{-1}$. Similarly, $\mu_\zeta$ is a
precision-weighted convex combination of the prior vector $\beta_0$ and the
empirical average of the variational posterior means $\mu_{1\:H}$ for
$\beta_{1\:H}$.

The updates for $\Upsilon$ and $\omega$ are similarly straightforward to
derive; we obtain
\begin{align}
  \omega & \leftarrow\ \nu + H \ , \label{eq:omega} \\
  \Upsilon & \leftarrow\ \left( S^{-1} +
      \sum_{h=1}^H \left( \Sigma_h +  (\mu_\zeta -
        \mu_h)(\mu_\zeta - \mu_h)^\top \right) +
      H \Sigma_\zeta\right)^{-1} \ . \label{eq:upsilon-update}
\end{align}
Notice that the solution~\myeqref{omega} for $\omega$ involves only the
constants $\nu$ and $H$. We compute $\omega$ once in advance, leaving it
unchanged during the variational optimization.

\section{An application of the delta method}
\label{apx:delta-method}

Let $f(v)$ be a function from $\mathbbm{R}^K$ to $\mathbbm{R}$. According
to the multivariate delta method for moments~\citep{BickelDoksum2007},
\begin{equation}
  \E f(V) \approx f(\E V) +
  \frac{1}{2} \tr \left[ \left( \hess{f(\E V)}{v} \right) \cov{V} \right] .
  \label{eq:delta-method}
\end{equation}
Consider the case
\begin{equation}
  f(v) = \log \left( 1^\top \exp(xv) \right) \ ,
\end{equation}
where $x$ is a $J \times K$ matrix whose rows are the vectors $x_j^\top$.
Let $V \sim \mathcal{N}_K(\mu, \Sigma)$, and restrict $\Sigma$ such that
$\Sigma = \diag\{ \exp(\sigma) \}$ for $\sigma \in \mathbb{R}^K$. We can
now rewrite~\myeqref{delta-method}:
\begin{equation}
  \E \log \left( 1^\top \exp(xV) \right) \approx
  \log \left( 1^\top \exp(x\mu) \right)
  + \frac{1}{2} \Theta(\mu)^\top \exp(\sigma) \ , \label{eq:delta-method-lse}
\end{equation}
where $\Theta(\mu)$ is the diagonal of the Hessian of $f$, evaluated at the
point $\mu$. Define $s \defeq 1^\top \exp(x \mu)$. Using matrix
differentials, it can be shown that
\begin{equation}
  \Theta(\mu) = s^{-1} (x \odot x)^\top \exp(x \mu)
  - s^{-2} \left( x^\top \exp(x \mu) \right) \odot
  \left( x^\top \exp(x \mu) \right) \ ,
\end{equation}
where $\odot$ denotes the Hadamard product.

To use the approximation~\myeqref{delta-method-lse} in an optimization over
$\mu$, we need to compute the gradient. The formula for $\Theta(\mu)$ makes
this a more extensive but still mechanical exercise in differentials. One
obtains
\begin{multline}
  \frac{\partial}{\partial \mu} =
  s^{-1} x^\top e^{x \mu} +
  \frac{1}{2} x^\top
  \left[
    \left(
      s^{-1} \diag\left\{e^{x \mu}\right\}
      - s^{-2} e^{x \mu} \left(e^{x \mu}\right)^\top
    \right) (x \odot x)
    \ + \right. \\
    \left. 2 \left(
      s^{-3}
      e^{x \mu}
      \left\{
        \left(x^\top e^{x \mu}\right) \odot \left(x^\top e^{x \mu}\right)
      \right\}^\top
      - s^{-2} \diag \left\{ e^{x \mu} \right\} x
      \diag \left\{ x^\top e^{x \mu} \right\}
    \right)
  \right] \exp(\sigma) \ .
\end{multline}


\section{A convexity result} \label{apx:convexity}

Let $a_1, \ldots, a_d$ be scalars, $c_1, \ldots, c_d$ be $n$-vectors, $p, r
> 0$, and $Q \succeq 0$. We show here that the function
\begin{equation}
f(B) = r \log \bigl| B B^T \bigr| - p \tr\left(Q B B^\top
\right)- \log \biggl(\sum_{j=1}^d \exp \left\{ a_j + c_j^\top B B^\top c_j
  \right\} \biggr) \label{eq:fB}
\end{equation}
is concave on the set of full-rank $n \times n$ matrices.

We argue that each of the three constituent terms, from left to right, is
concave. The second differential of $g(B) = r \log \bigl| B B^T \bigr|$ is
\begin{equation}
  \mrd^2 g = \mrd \tr\left\{ 2 r B^{-1} \mrd B\right\}
  = \tr\left\{ -2r \left[ B^{-1} (\mrd B) \right]^2 \right\} \ .
\end{equation}
By Theorem~10.6.1 of~\citep{MagnusNeudecker2007}, the Hessian of $g$ is \\
$-2 r K_n \left( B^{-\top} \otimes B^{-1} \right)$, where $K_n$ is the
order-$n$ commutation matrix and $\otimes$ denotes the Kronecker product.
We now show that $K_n \left( B^{-\top} \otimes B^{-1} \right)$ is (matrix)
positive-definite.
\begin{align}
  (\mrd \vecop X)^\top K_n \left( B^{-\top} \otimes B^{-1} \right) (\mrd
  \vecop X) & =
  (\mrd \vecop X)^\top K_n \vecop \left\{ B^{-1} (\mrd X) B^{-1} \right\}
  \label{eq:step1} \\
  & = (\vecop \mrd X)^\top \vecop \left\{ B^{-\top} (\mrd X)^\top
    B^{-\top} \right\} \\
  & = \tr \left\{ \left( B^{-1} \mrd X \right)^2 \right\} \geq 0 \ .
\end{align}
Equation~\myeqref{step1} follows from the well-known fact that $\vecop ABC
= (C^\top \otimes A) \vecop B$. Thus, the Hessian of $g$ is negative
definite, and $r \log \bigl| B B^T \bigr|$ is concave.

Concavity of the middle term in~\myeqref{fB} follows in the usual way from
the univariate convexity of the function
\begin{equation}
  g(t) \defsym \tr \left( Q (M + tP)(M + tP)^\top \right)
  = \sum_{i=1}^n (m_i + t p_i)^\top Q (m_i + t p_i)
\end{equation}
for fixed matrices $M$ and $P$, with columns $m_i$ and $p_i$. To see that
the rightmost term in~\myeqref{fB} is concave, define
\[
g_j(t) \defsym a_j + c_j^\top (M + tQ) (M + tQ)^\top c_j
\]
for $j = 1,\ldots,d$ and fixed matrices $M$ and $Q$. Each $g_j$ is convex
in $t$, and the rightmost term in~\myeqref{fB} is (minus) the log-sum-exp
function composed with the $g_j$'s. Concavity of this term in $t$, and
hence in $B$, follows from~\citep[][p. 86]{BoydVandenberghe2004}.

\begin{singlespace}
\bibliography{../refs/braun_refs}
\end{singlespace}

\end{document}